\documentclass[
]{ceurart}

\usepackage{listings}
\begin{document}

\copyrightyear{2025}
\copyrightclause{Copyright for this paper by its authors.
  Use permitted under Creative Commons License Attribution 4.0
  International (CC BY 4.0).}

\conference{First Workshop on Large Language Models For Generative Software Engineering (LLM4SE 2025)}

\title{Optimizing Retrieval Augmented Generation for
Object Constraint Language}

\author[1]{Kevin Chenhao Li}[%
email=kevinchenhao.li@tum.de,
]
\cormark[1]

\author[1]{Vahid Zolfaghari}[%
email=v.zolfaghari@tum.de,
]

\author[1]{Nenad Petrovic}[%
email=nenad.petrovic@tum.de,
]

\author[1]{Fengjunjie Pan}[%
email=f.pan@tum.de,
]

\author[1]{\textcolor{black}{Prof.} Alois Knoll}[%
email=k@tum.de,
]

\address[1]{Technical University of Munich (TUM), Arcisstraße 21
D-80333 Munich, Germany}

\cortext[1]{Corresponding author.}

\begin{abstract}
  The Object Constraint Language (OCL) is essential for defining precise constraints within Model-Based Systems Engineering (MBSE). However, manually writing OCL rules is complex and time-consuming. This study explores the optimization of Retrieval-Augmented Generation (RAG) for automating OCL rule generation, focusing on the impact of different retrieval strategies. We evaluate three retrieval approaches—BM25 (lexical-based), BERT-based (semantic retrieval), and SPLADE (sparse-vector retrieval)—analyzing their effectiveness in providing relevant context for a large language model.

To further assess our approach, we compare and benchmark our retrieval-optimized generation results against PathOCL, a state-of-the-art graph-based method. We directly compare BM25, BERT, and SPLADE retrieval methods with PathOCL to understand how different retrieval methods perform for a unified evaluation framework. Our experimental results, focusing on retrieval-augmented generation, indicate that while retrieval can enhance generation accuracy, its effectiveness depends on the retrieval method and the number of retrieved chunks (k). BM25 underperforms the baseline, whereas semantic approaches (BERT and SPLADE) achieve better results, with SPLADE performing best at lower k values. However, excessive retrieval with high k parameter can lead to retrieving irrelevant chunks which degrades model performance. Our findings highlight the importance of optimizing retrieval configurations to balance context relevance and output consistency. This research provides insights into improving OCL rule generation using RAG and underscores the need for tailoring retrieval.
\end{abstract}


\maketitle

\section{Introduction}

Object Constraint Language \cite{b1} plays an important role in Model-Based Systems Engineering (MBSE) by enabling precise constraint definition within meta-models. OCL is used to ensure the integrity of system designs by specifying conditions that must be held within a model. It is widely used in Unified Modeling Language (UML) \cite{b2} and Eclipse Modeling Framework (EMF) \cite{b3} by defining invariants, preconditions, postconditions, and derived attributes, which cannot be specified by the model itself and thereby enhancing model expressiveness.

However, manually writing OCL rules is complex and time-consuming, requiring a deep understanding of both the system model and OCL syntax. Natural language is often the starting point for defining system constraints, such as those given in the requirements and specifications of the system. This makes an automated approach that translates natural language specifications into OCL rules highly attractive and could significantly improve efficiency and accessibility.

Recent advances in Large Language Models (LLMs) have revolutionized automated code and rule generation. Models such as GPT-4 \cite{b4}, DeepSeek \cite{b5}, and Meta-Llama-3 \cite{b6} have demonstrated remarkable capabilities in understanding and generating structured text, including programming languages and domain-specific rule sets. These models leverage pre-trained knowledge from large text corpora, enabling them to generalize across different programming languages, formal notations, and syntactic structures. In particular, LLMs have been thoroughly applied in natural language-to-code translation \cite{b7} \cite{b8}, showing strong performance in converting natural language instructions into executable code in languages such as Python, Java, and SQL. Similarly, they can be adapted to translate natural language specifications into OCL rules \cite{b9} \cite{b10}, reducing the manual effort required by human engineers. However, LLMs often struggle with domain-specific knowledge \cite{b11}, especially when dealing with complex specifications and extensive meta-models. A meta-model defines the structure and rules for how models are built in MBSE. It includes elements such as classes, associations, enumerations, and attributes. This is where RAG becomes essential. Large Language Models have a limited context window, making it challenging to include large and complex meta-models entirely in a prompt. RAG allows relevant parts of the meta-model to be retrieved and injected dynamically, helping the LLM generate precise rules even when the whole meta-model does not fit into the context window.

\begin{figure}[htbp]
\centerline{\includegraphics[width=0.4\linewidth]{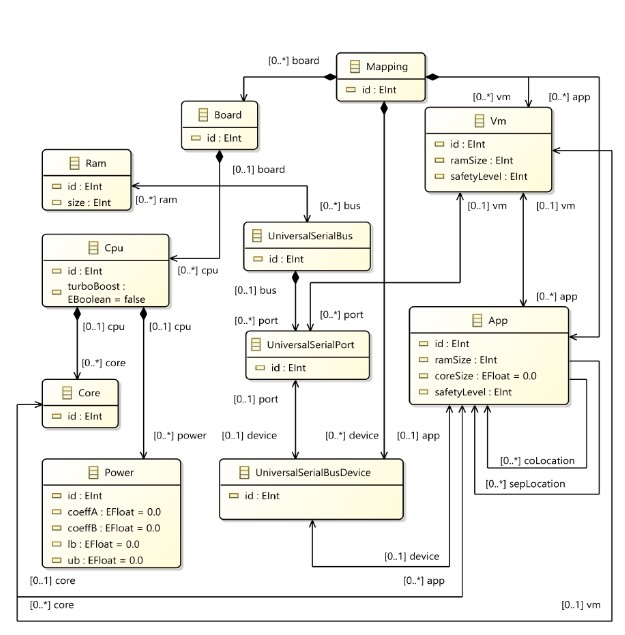}}
\caption{Example meta-model from \cite{b9}}
\label{fig:ExampleMetamodel}
\end{figure}

Retrieval-Augmented Generation is an approach that enhances LLMs by integrating external knowledge retrieval into the generation process \cite{b12} \cite{b13}. This idea originated from the question answering domain \cite{b14}. Instead of relying solely on a model’s pre-trained knowledge, RAG retrieves relevant information from an external knowledge base and incorporates it into the model’s input before generating the final output. This method has shown promise in improving accuracy, reducing hallucinations \cite{b15}, and ensuring that generated content aligns with the domain \cite{b13}. Since OCL rules are tightly coupled with the underlying meta-model structure, a standard LLM may not have sufficient context to generate the correct rule. RAG allows us to retrieve relevant meta-model elements (e.g., classes, associations, enumerations) from a retrievable knowledge base and include them in the input prompt. The desired result is that RAG helps the LLM generate rules that adhere to proper OCL syntax and semantics. However, optimizing retrieval strategies for OCL generation has not been extensively studied, particularly in the context of balancing retrieval efficiency and generation accuracy.

While fine-tuning can adapt LLMs to specific tasks, such as OCL rule generation, it is resource-intensive and may not generalize well across new or unseen meta-models. Recent research has increasingly highlighted the advantages of Retrieval-Augmented Generation over fine-tuning for knowledge injection in large language models. RAG consistently outperforms fine-tuning across multiple datasets, even when dealing with previously known and entirely new knowledge. Fine-tuning struggles with learning new factual information and requires extensive training data \cite{b16}. Similarly, another study \cite{b17} found that RAG is more effective in handling less popular or low-frequency knowledge, including domain-specific knowledge. The study emphasizes that while smaller language models may still benefit from fine-tuning, larger models gain little additional advantage. Additionally, fine-tuning remains resource-intensive. Combining RAG with fine-tuning can lead to further improvements to performance in specialized domains as shown in \cite{b18}.

Current research in OCL rule generation has focused on other approaches, such as fine-tuning \cite{b9}. This study uses RAG but does not evaluate the impact of it or experiment with different retrieval configurations. Other current work about OCL rule generation using LLMs include \cite{b10} and \cite{b23}, which either fully inject the meta-model without retrieval or use a path-based approach. However, the impact of retrieval-based methods such as RAG on OCL rule generation remains underexplored. This gap in the literature highlights the necessity of further investigating retrieval-based strategies to enhance the accuracy and efficiency of OCL rule generation. By evaluating different retrieval configurations and assessing their impact, this study aims to contribute a novel perspective to the field.

This study explores the optimization of RAG for OCL rule generation, focusing on different retrieval approaches to enhance model performance. We investigate traditional lexical-based retrieval (BM25) \cite{b19}, semantic dense-vector retrieval using transformer-based models (BERT) \cite{b20}, and semantic sparse-vector retrieval (SPLADE) \cite{b21}. BM25 is a keyword-matching method that scores documents based on term frequency and inverse document frequency. BERT-based retrieval uses contextual embeddings to capture semantic similarity between queries and documents. SPLADE, on the other hand, creates sparse representations of text using learned term expansions, combining term matching and the ability to find nearest neighbors. While these methods have been extensively studied in general NLP tasks, their effectiveness in a RAG approach to OCL rule generation remains unexplored. This study bridges that gap by systematically evaluating retrieval strategies and their impact on generation accuracy. We aim to identify optimal configurations for improving OCL constraint generation. We compare our optimal configurations against other state-of-the-art methods for generating OCL constraints. Our evaluation employs quantitative metrics such as Cosine Similarity and Euclidean Distance to assess model output quality. This research contributes to the growing body of work in domain-adapted LLM applications and provides insights into improving automated OCL constraint generation. 

The results demonstrate that while retrieval strategies can enhance generation quality, they must be carefully tuned to avoid performance degradation due to excessive or irrelevant retrieved information. Our findings highlight the importance of selecting an appropriate retrieval method and the optimal number of retrieved chunks ($k$) to maximize performance. This study provides insights into the impact of different retrieval techniques and lays the foundation for future improvements in automated OCL rule generation.

\section{Methodology}

\subsection{Pipeline}

\begin{figure}[htbp]
\centerline{\includegraphics[width=0.8\linewidth]{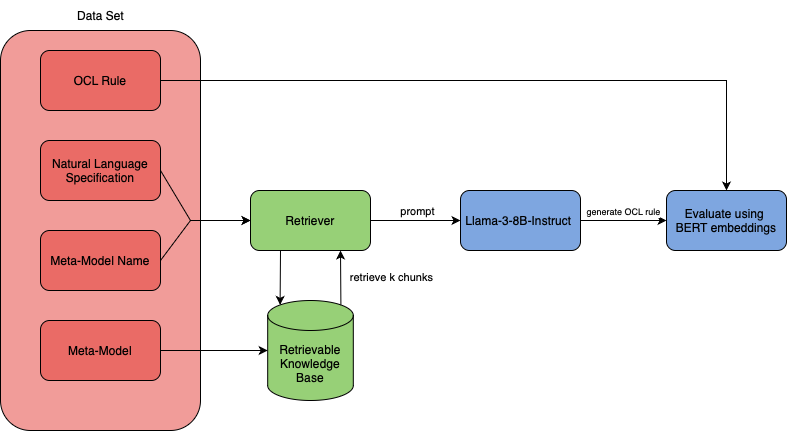}}
\caption{Retrieval Augmented Generation Pipeline}
\label{fig:pipeline}
\end{figure}

We first give a brief overview of the entire pipeline in this section before going into detail for every step of the pipeline in the subsequent sections.
Our pipeline takes as input a natural language specification of an OCL rule and the associated name of the meta-model that we want to generate the OCL rule for. We use both parts of the input in the retrieval stage to find relevant chunks of the meta-model. These retrieved chunks are then incorporated into a prompt alongside the natural language specification and given to a Large Language Model. We then compare the output of the LLM with the actual OCL rule to determine the quality of our output. This workflow can be seen in Figure \ref{fig:pipeline}.

\subsection{Environment Setup}

The experiment was carried out using the free version of Google Colab, using the T4 GPU for access to computing resources. Necessary dependencies were installed, including transformers, bitsandbytes, flash-attn, and pyngrok. In addition, a Hugging Face authentication token was configured to facilitate secure access to the model repository and data set.

The Meta-Llama-3-8B-Instruct model was selected and loaded using the transformers library. The selection of LLaMA-3-8B-Instruct model was due to its performance, accessibility, and resource efficiency. Its 8 billion parameter size is significantly smaller than models like GPT-4 enabling us to run it without the need to pay for API calls. The role of the model was set to "system" and we further limited the maximum length of the output to 1024 tokens. Since we do not further train the pre-trained model and are only interested in the evaluation of the output, we set the \texttt{do\_sample} flag of the Meta-Llama-3-8B-Instruct model to false to disable random sampling and use greedy decoding to improve reproducibility. This ensures that any observed variance in outputs is attributable to changes in retrieval context rather than sampling noise.

A web service was implemented using Flask, a lightweight Python web framework. An REST API was created to handle incoming requests, process inputs, and generate responses from the LLM. The API was structured to receive user queries, pass them through the model, and return the generated text output.

\subsection{Data set}

We used the data set collected by Pan et al. \cite{b9} \footnote{This data set can be found at https://huggingface.co/datasets/fpan/text-to-ocl-from-ecore.}. Each sample in the data set consists of:

\begin{itemize}
    \item OCL rule
    \item Natural language specification of the OCL rule
    \item Name of meta-model
    \item Textual description of meta-model given in PlantUML format
\end{itemize}

The dataset was preprocessed by segmenting the PlantUML strings so that each chunk contained only a single class, enumeration, or association. This ensured that each chunk is semantically complete and is the smallest atomic unit that cannot be further divided without losing meaning. We implemented the chunking via stop words, where a chunk is considered to end if we encounter either one of "class", "enum" or "association". During this pre-processing step certain characters like tabs and unnecessary formatting like line breaks were removed, which were present in the original data set. This resulted in a total of 3595 unique chunks over the entire dataset. Using the chunks we then built our external knowledge base, where for each meta-model, uniquely identified by its name, we have a collection of chunks representing the whole meta-model.

To evaluate retrieval impact, we filtered the dataset such that only hard samples were considered. Hard samples were defined as instances where the number of chunks for the meta-model exceeded 50, making retrieval non-trivial and requiring a retriever to select a subset of chunks to use in the context of the prompt. From that filtered dataset we then randomly sampled 72 instances for our evaluation, which corresponds to one-fifth of the size of the original dataset.

\subsection{Retrieval}

To enhance the accuracy and relevance of generated responses, a RAG pipeline was integrated. We built the external knowledge base from the textual PlantUML provided in the data set as described before. The RAG pipeline retrieves relevant chunks from the knowledge base before passing them as context to the language model based on the natural language specification. We applied two filtering conditions for relevance: (a) chunks must belong to the same meta-model as the input sample, and (b) chunks must score higher than others based on similarity. Both conditions must be satisfied to be selected.

Different retrieval approaches were evaluated, including lexical-based approaches in BM25-based retrieval, and transformer-based retrieval models based on dense and sparse vectors such as BERT and SPLADE. For all retrieval models, we evaluated them using top-k retrieval, where the top-k chunks regarding the retrieval score with the natural language specification were given as context to the LLM. For each retrieval model, we evaluated them with $k$ set to 10, 20, 30, 40, and 50. We also evaluated the intrinsic performance of the LLM regarding OCL rule generation using no retrieval.

For our BM25-based retriever, we tokenized the natural language specification and used the result as our query for the BM25 algorithm. For the transformer-based, approaches we compared the embeddings between the natural language specification and the chunks of the meta-model and selected the top-k most similar chunks using cosine similarity. We used the cosine similarity implementation from the scikit-learn library.

\subsection{Generation}

\begin{figure}
    \begin{lstlisting}[basicstyle=\ttfamily\small, breaklines=true, caption={Prompt used in the pipeline}, label={lst:prompt}]
You are given a meta-model with information about classes, associations and their attributes.
You are also given a natural language specification.
Your task is to generate an OCL (Object Constraint Language) constraint for this specification 
and based on the meta-model.
Do not provide any explanations or additional text.
The meta-model information is: {retrieved chunks}
The natural language specification is: {specification}
\end{lstlisting}
\end{figure}

To generate the output we used a prompt that was slightly adapted from \cite{b9}. We added "Do not provide any explanations or additional text." to discourage the LLM from outputting lengthy responses that negatively impact its performance regarding our automated metrics. The final prompt template is shown in Listing \ref{lst:prompt}

\subsection{Evaluation}

Various configurations of the RAG pipeline were tested to assess their impact on response quality. This included differing the parameters such as the number of retrieved chunks, and the embedding model selection.

The evaluation was conducted using automated quantitative metrics comparing the generated model output to the actual OCL rule as given in the data set. We use cosine similarity and Euclidean distance as implemented by the \texttt{scikit-learn} library and based on BERT embeddings as our evaluation metrics. The choice of the BERT embedding model is due to its open-source nature. The proposed evaluation methodology provides an efficient and scalable way to measure the output quality, allowing us to compare a large number of retriever configurations against each other.

\section{Results}
We will first present the results for each retrieval approach and then compare the results across different retrieval approaches. The results are based on the random subset of the filtered data set by Pan et al. \cite{b9}. Given that the best possible cosine similarity and the best possible Euclidean distance to the original OCL rule are 1 and 0 respectively, the y-axis is inverted when displaying Euclidean distances. We consider variances close to 0 as more desirable, as they represent output and performance consistency. We also examined the performance when disregarding the worst 10\% of generated samples to determine the generation output quality without extreme outliers such as hallucinations or outputs disregarding the instruction not to explain the answer. Trimmed mean refers to the mean calculated after removing the worst 10\% of samples based on similarity score. The results are rounded to the 4th decimal place to improve readability. We abbreviate Cosine Similarity with CS and Euclidean Distance with ED.

Our results highlight the impact of different retrieval approaches on the performance of the RAG pipeline for OCL rule generation. The baseline results as seen in Table \ref{tab:BM25}, where no retrieval was applied (k=0), indicate that the language model alone achieves a relatively high cosine similarity (0.9338) but still leaves room for improvement through the integration of retrieval strategies. This result indicates that the intrinsic knowledge in the domain of OCL rules or at least the ability to generate similar rules gives us a strong baseline in regards to our semantic similarity metrics.

\subsection{Effectiveness of Different Retrieval Approaches}
We evaluated BM25 as a lexical retriever using cosine similarity and Euclidean distance to measure performance. The results are shown in Table \ref{tab:BM25} and compared with the no-retrieval baseline. Table \ref{tab:BERT} presents the results for BERT-based retrieval, which uses dense semantic embeddings for chunk selection. SPLADE-based retrieval results are shown in Table \ref{tab:SPLADE}. SPLADE uses sparse semantic representations, optimized for balancing lexical precision with semantic flexibility. 

Comparing the retrieval methods, the BM25-based retriever (Table \ref{tab:BM25}) exhibited a decline in performance compared to the baseline, particularly at higher values of $k$, suggesting that lexical retrieval alone is insufficient for effective context selection. The best performance for BM25 was observed at $k = 30$ with a mean cosine similarity of 0.9292, which still underperforms our baseline across all metrics. Generally, the variance in cosine similarity and Euclidean distance for BM25-based retrieval was higher than in other retrieval approaches, indicating inconsistent retrieval performance as seen in Fig. \ref{fig:BM25_COSSIM} and Fig. \ref{fig:BM25_EUCDIS}. We have marked the best-performing model in each table by using bold numbers.

\begin{table*}
  \caption{BM25-based retriever}
  \label{tab:BM25}
  \begin{tabular}{ccccccl}
    \toprule
    \textbf{Metric}& \textbf{\textit{k = 10}}& \textbf{\textit{k = 20}}& \textbf{\textit{k = 30}}& \textbf{\textit{k = 40}}& \textbf{\textit{k = 50}} & \textbf{\textit{k = 0}} (Baseline) \\
    \midrule
    Mean CS  & 0.9231 & 0.9208 & \textbf{0.9292} & 0.9212 &  0.9133 & 0.9338 \\
    Variance CS  & 0.0028 & 0.0023 & \textbf{0.0021} & 0.0042 & 0.0064 & 0.0022 \\
    Trimmed Mean CS  & 0.9366 & 0.9321 & \textbf{0.9403} & 0.9364 & 0.9348 & 0.946 \\
    \midrule
    Mean ED  & 5.434 & 5.5972 & \textbf{5.2179} & 5.4504 & 5.6474 & 5.0682 \\
    Variance ED  & 3.5722 & \textbf{3.2460} & 3.2735 & 4.6489 & 6.4171 & 3.2026 \\
    Trimmed Mean ED  & 5.0244 & 5.2199 & \textbf{4.8489} & 5.0026 & 5.0397 & 4.6396 \\
  \bottomrule
\end{tabular}
\end{table*}

Lexical approaches rely heavily on exact matching.  We hypothesize that this approach struggles because there is no guarantee that our natural language description uses the exact terms that are present in the relevant chunks. The results point to no context being better than misleading or incomplete context, which is supported by current research \cite{b22}.

Presented in Table \ref{tab:BERT}, the BERT-based retrieval model demonstrated more stable performance across different values of $k$, achieving a mean euclidean distance of 5.0418 at $k=50$, which is slightly better than the baseline. This suggests that semantic similarity-based retrieval can contribute positively to the overall generation quality. Additionally, the lower variance in cosine similarity and Euclidean distance for BERT-based retrieval, as illustrated in Fig. \ref{fig:BERT_COSSIM} and Fig. \ref{fig:BERT_EUCDIS}, included in the appendix, suggests a more consistent performance across different samples. When looking at the performance with 10\% of the worst samples removed, we observe that the model with $k=50$ is no longer the best-performing one. This indicates that when disregarding consistency as defined by the ability to limit outliers, a slightly lower value for $k$ might provide better performance.

\begin{table*}
  \caption{BERT-based retriever}
  \label{tab:BERT}
  \begin{tabular}{ccccccl}
    \toprule
    \textbf{Metric}& \textbf{\textit{k = 10}}& \textbf{\textit{k = 20}}& \textbf{\textit{k = 30}}& \textbf{\textit{k = 40}}& \textbf{\textit{k = 50}} & \textbf{\textit{k = 0}} (Baseline)\\
    \midrule
    Mean CS  & 0.9265 & 0.9263 & 0.9254 &0.9259 & \textbf{0.9334} & 0.9338  \\
    Variance CS  & 0.0045 & 0.0036 & 0.0064 & 0.0050  & \textbf{0.0017} & 0.0022 \\
    Trimmed Mean CS  & 0.9435 & 0.9415 & 0.9437 & \textbf{0.9438} & 0.9425 & 0.946  \\
    \midrule
    Mean ED  & 5.2026 & 5.2436 & 5.163 &  5.2477 & \textbf{5.0418} & 5.0682  \\
    Variance ED  & 4.8764 & 4.3763 & 6.2543 & 5.1349 & \textbf{3.2301} & 3.2026  \\
    Trimmed Mean ED  & 4.6916 & 4.7734 & \textbf{4.6247} & 4.7097 & 4.7065 & 4.6396 \\
  \bottomrule
\end{tabular}
\end{table*}

\begin{figure}[htbp]
\centerline{\includegraphics[width=0.9\linewidth]{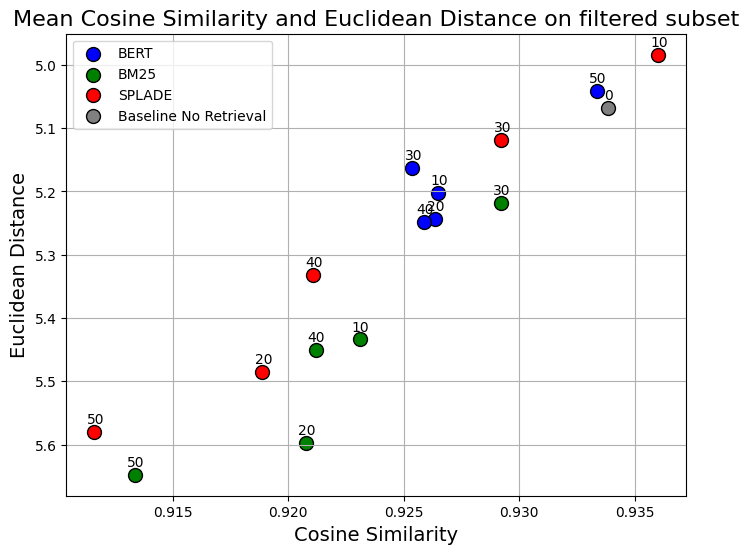}}
\caption{Comparison of Retriever Models}
\label{fig:COMPARISON}
\end{figure}

The SPLADE-based retriever produced mixed results, as shown in Table \ref{tab:SPLADE}, showing relatively high cosine similarity at lower values of $k=10$ but experiencing a strong decline in performance as $k$ increased. Notably, SPLADE at $k=10$ outperformed all other retrieval approaches, including the baseline and the best-performing BERT-based retriever (Fig. \ref{fig:COMPARISON}), suggesting that sparse-vector retrieval models may be particularly beneficial when selecting a limited number of relevant chunks. We hypothesize that SPLADE can leverage exact matching and synonyms to find all relevant chunks quickly. The performance then degrades when we increase k as little to no additional relevant chunks are included. Similar to BERT-based retrieval, our SPLADE-based has a stronger performance for $k = 30$ when disregarding outliers. Once again this observation is in part due to the advantage of higher consistency across outputs for $k = 10$, but suggests that a different value for $k$ might be more beneficial when removing outliers. While the absolute improvement in mean cosine similarity appears numerically small, the difference can be semantically meaningful for domain-specific tasks like this one.

\begin{table*}
  \caption{SPLADE-based retriever}
  \label{tab:SPLADE}
  \begin{tabular}{ccccccl}
    \toprule
    \textbf{Metric}& \textbf{\textit{k = 10}}& \textbf{\textit{k = 20}}& \textbf{\textit{k = 30}}& \textbf{\textit{k = 40}}& \textbf{\textit{k = 50}} & \textbf{\textit{k = 0}} (Baseline)\\
    \midrule
    Mean CS  & \textbf{0.9360} & 0.9189 & 0.9292 & 0.9211 & 0.9116 & 0.9338 \\
    Variance CS  & \textbf{0.0016} & 0.0049 & 0.0042 & 0.0068 & 0.0100 & 0.0022\\
    Trimmed Mean CS  & 0.9453 & 0.9370 & \textbf{0.9461} & 0.9423 & 0.9408 & 0.946\\
    \midrule
    Mean ED  & \textbf{4.9842} & 5.4851 & 5.1187 & 5.3317 & 5.5794 & 5.0682\\
    Variance ED  & \textbf{2.7181} & 5.3273 & 4.6761 & 6.3177 & 7.9934 & 3.2026\\
    Trimmed Mean ED  & 4.6473 & 4.9433 & \textbf{4.5949} & 4.7300 & 4.8053 & 4.6396 \\
  \bottomrule
\end{tabular}
\end{table*}

\subsection{Impact of k}

We also analyzed how varying the number of retrieved chunks as determined by the parameter $k$ affects model performance. Interestingly, increasing $k$ does not always lead to improved results. For the BM25 and SPLADE retrievers, performance fluctuated as $k$ increased, suggesting that excessive retrieval may introduce irrelevant or redundant information, as shown in Table \ref{tab:BM25} and Table \ref{tab:SPLADE}. We speculate that excessive retrieval (higher $k$) can introduce noise, leading to lower similarity scores. Conversely, BERT-based retrieval demonstrated relatively stable performance across different k values, with its best performance occurring at $k = 50$, as seen in Table \ref{tab:BERT}.  Our results suggest that optimizing the retrieval step by carefully selecting an appropriate $k$ is crucial to maximizing the benefits of retrieval-augmented generation in the OCL domain, and blindly increasing $k$ can degrade the model’s effectiveness. 

Furthermore, the analysis of variance across both evaluation metrics highlights the importance of retrieval stability. Models with lower variance in cosine similarity (such as the BERT-based approach at $k=50$) tend to be more reliable in producing overall high-quality outputs, illustrated in Fig \ref{fig:BERT_COSSIM} and \ref{fig:BERT_EUCDIS}. Inversely, higher variance in the BM25-based and SPLADE-based retrieval approaches at almost all $k$ values suggests inconsistency, potentially due to the inclusion of less relevant chunks in the retrieved context, as illustrated in Fig. \ref{fig:BM25_COSSIM}, \ref{fig:BERT_EUCDIS}, \ref{fig:SPLADE_COSSIM}, and \ref{fig:SPLADE_EUCDIS}. These results indicate that changing the retriever model and parameter $k$ can have a positive impact on hallucination and reducing outliers, but cannot fully remove them. Depending on the needs and goals of a potential end user, a trade-off between model consistency and model output quality in a large percentage of cases needs to be considered.

\subsection{Comparison with PathOCL}
PathOCL is a novel path-based prompt augmentation method proposed by Abukhalaf et al. \cite{b23}. The approach constructs a graph based on the PlantUML of the meta-model, where each class is represented as a node of the graph and associations are directed edges. The direction of the graph is dependent on the type of the association and its direction in the meta-model. PathOCL extracts all simple paths through the graph and ranks them based on their similarity to the natural language specification using either Jaccard or cosine similarity. For a natural language specification of an OCL constraint the approach extracts the UML elements using POS-tagging and then ranks all simple paths in the graph based on either the jaccard similarity or cosine similarity between the extracted elements and the node names along the path. The most relevant paths are then included in the prompt to help the LLM generate the correct OCL constraint.

Although their approach aims to retrieve relevant classes in the face of limited context size, the dataset they use to evaluate their approach only consists of 15 UML models, where the largest model only contains 11 classes and 10 associations. Furthermore the PlantUML files of their dataset did not contain interfaces, enum, composition relations or aggregation relations. We evaluated the PathOCL method on the larger and harder dataset provided by \cite{b9}. One issue that was raised is the runtime on extremely large meta-models. Given the length of some of the meta-models, having upwards of 100 classes and 300 associations makes it almost impossible to perform the method as outlined, as it uses a brute-force approach to compute all simple paths in the graph, which has a runtime complexity of at least $O(n!)$. 

We thus decided to do the opposite as mentioned in the methodology section and filter our dataset by only considering samples where the number of chunks in the meta-model is less than 100 and randomly sampled 72 instances. We evaluated PathOCL under both Jaccard and cosine similarity configurations with different values of $k=(1, 3, 5)$. As shown in Table \ref{tab:PathOCLBest}, our SPLADE-based method at $k=10$ outperformed PathOCL across both cosine similarity and Euclidean distance. Additional comparative PathOCL results for varying $k$ values and similarity measures are presented in Tables \ref{tab:PathOCLJaccard} and \ref{tab:PathOCLCosine}. Our results indicate that sparse-vector retrieval provides a significant increase in performance against the no retrieval baseline. Counterintuitively the PathOCL approach is outperformed by the baseline as seen in Table \ref{tab:PathOCLBest}. These results suggest that semantic retrieval strategies scale better on complex datasets.

\begin{table*}
  \caption{Performance on subset with maximum length for meta-models}
  \label{tab:PathOCLBest}
  \begin{tabular}{cccl}
    \toprule
    \textbf{Metric}&\textbf{PathOCL (Jaccard \textit{k=1})}&\textbf{Baseline}&\textbf{SPLADE \textit{k=10}}\\
    \midrule
    Mean CS  & 0.9251 & 0.9280 & 0.9328\\
    Mean ED  & 5.3356 & 5.2775 & 4.9725\\
  \bottomrule
\end{tabular}
\end{table*}

\begin{table*}
  \caption{PathOCL performance with Jaccard Similarity}
  \label{tab:PathOCLJaccard}
  \begin{tabular}{cccl}
    \toprule
    \textbf{Metric}&\textbf{Jaccard \textit{k=1}}&\textbf{Jaccard \textit{k=3}}&\textbf{Jaccard \textit{k=5}}\\
    \midrule
    Mean CS  & 0.9251 & 0.9066 & 0.8986\\
    Mean ED  & 5.3356 & 5.8805 & 6.2185\\
  \bottomrule
\end{tabular}
\end{table*}

\begin{table*}
  \caption{PathOCL performance with Cosine Similarity}
  \label{tab:PathOCLCosine}
  \begin{tabular}{cccl}
    \toprule
    \textbf{Metric}&\textbf{Cosine \textit{k=1}}&\textbf{Cosine \textit{k=3}}&\textbf{Cosine \textit{k=5}}\\
    \midrule
    Mean CS  & 0.9204 & 0.8957 & 0.9030\\
    Mean ED  & 5.4709 & 6.1700 & 6.0020 \\
  \bottomrule
\end{tabular}
\end{table*}

\subsection{Limitations and Future Work}

While our study demonstrates the possible benefits of retrieval-based approaches, it is not without limitations. First, our evaluation was conducted on a relatively small filtered subset of the dataset, which may not generalize to all OCL rule generation scenarios. While these metrics provide an initial assessment of textual similarity and closeness, they do not capture functional correctness. To complement our automated metrics, future work could add validation of whether generated OCL rules conform to formal OCL syntax, human expert review, and in-depth error analysis.

Moreover, while we evaluated diverse retrieval approaches in our experiments, further research is needed to explore more advanced retrieval techniques, such as hybrid approaches like multi-stage retrieval. Fine-tuning retrieval models specifically for OCL constraints may also yield additional performance gains over our base models. Another underexplored way to improve the generation of OCL rules based on natural language specifications could be refining the chunking strategy to ensure that retrieved information is both concise and semantically rich, for example by grouping chunks that are semantically connected. Instead of just retrieving meta-model chunks, it could also be beneficial to retrieve appropriate best practices and OCL examples, to leverage in context learning.

\section{Conclusion}

Our study investigated the impact of different retrieval strategies on the performance of a Retrieval-Augmented Generation pipeline for generating OCL rules. We evaluated three retrieval methods, BM25, BERT-based retrieval, and SPLADE-based retrieval, analyzing their effectiveness in providing relevant context for a large language model.

Our findings indicate that while retrieval can enhance generation accuracy, its effectiveness is highly dependent on the retrieval method and the number of retrieved chunks $k$. BM25-based retrieval underperformed the baseline, likely due to its reliance on exact term matching, which may not always align with natural language specifications. In contrast, semantic retrieval approaches such as BERT and SPLADE provided better performance, with SPLADE achieving the best results at lower $k$ values but degrading at higher $k$ values due to the inclusion of less relevant context.

A key takeaway is that blindly increasing $k$ does not always yield better results. Instead, an optimal retriever-dependent balance must be struck to avoid retrieval-induced noise while ensuring sufficient context for the generation model. Additionally, we observed that retrieval approaches with lower variance in performance provide more reliable and overall better results, which may be preferable in practical applications where consistency is crucial.

Additionally, our comparison with the PathOCL method highlights that our RAG-based approach, particularly SPLADE with $k=10$, outperforms graph-based path selection, especially on larger and more complex meta-models.

\begin{acknowledgments}
This research was funded by the Federal Ministry of Education and Research of Germany (BMBF) as part of the CeCaS project, FKZ: 16ME0800K.
\end{acknowledgments}

\section*{Declaration on Generative AI}
 During the preparation of this work, the author(s) used Grammarly in order to: Grammar and spelling check. After using these tool(s)/service(s), the author(s) reviewed and edited the content as needed and take(s) full responsibility for the publication’s content. 

\newpage

\bibliography{references}

\newpage

\appendix

\section{Plots}
\begin{figure}[htbp]
\centerline{\includegraphics[width=0.8\linewidth]{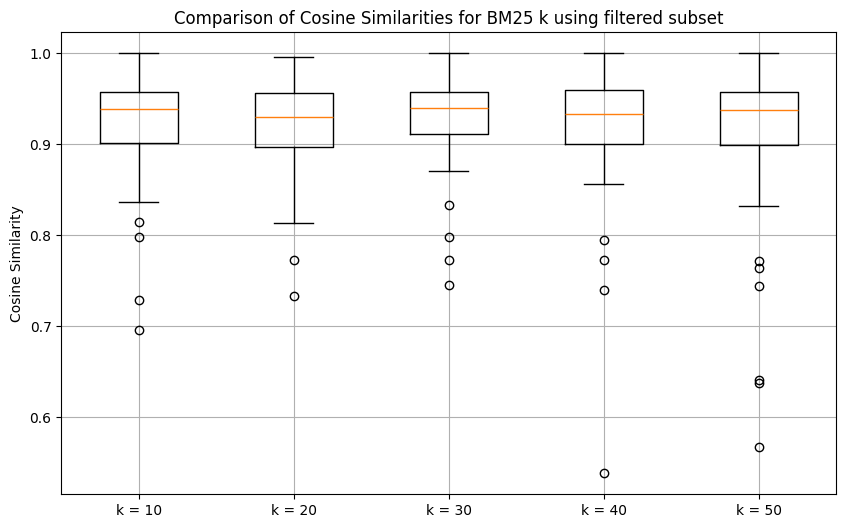}}
\caption{Boxplot of Cosine Similarities BM25}
\label{fig:BM25_COSSIM}
\end{figure}

\begin{figure}[htbp]
\centerline{\includegraphics[width=0.8\linewidth]{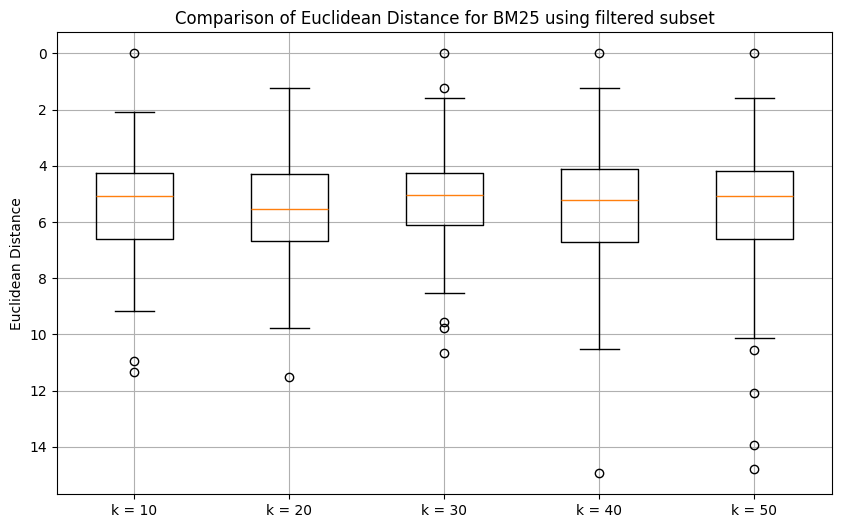}}
\caption{Boxplot of Euclidean Distances BM25}
\label{fig:BM25_EUCDIS}
\end{figure}

\begin{figure}[htbp]
\centerline{\includegraphics[width=0.8\linewidth]{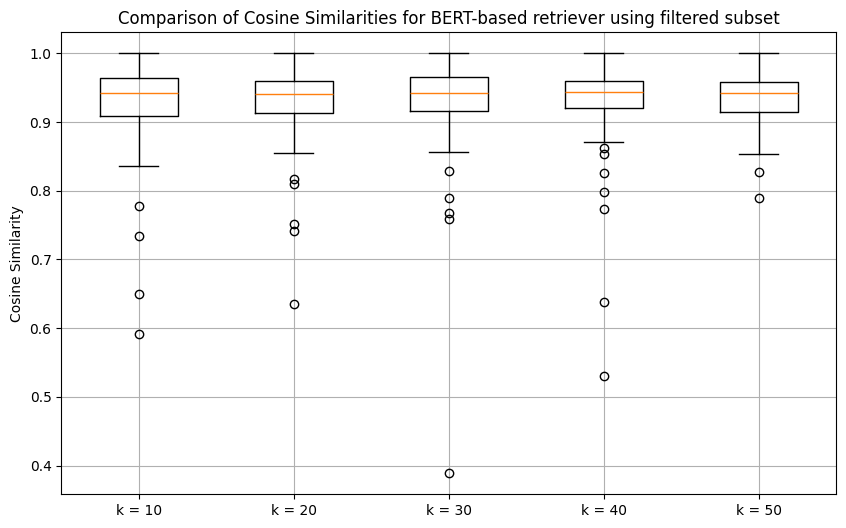}}
\caption{Boxplot of Cosine Similarities BERT}
\label{fig:BERT_COSSIM}
\end{figure}

\begin{figure}[htbp]
\centerline{\includegraphics[width=0.8\linewidth]{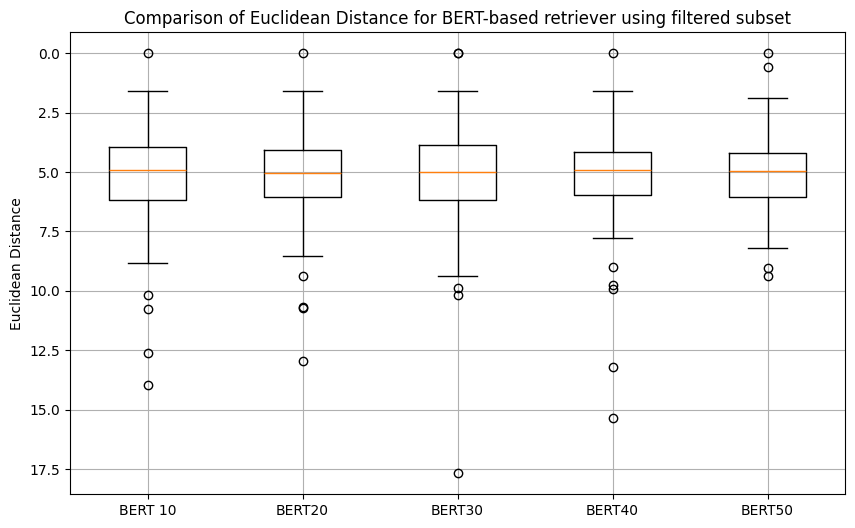}}
\caption{Boxplot of Euclidean Distances BERT}
\label{fig:BERT_EUCDIS}
\end{figure}

\begin{figure}[htbp]
\centerline{\includegraphics[width=0.8\linewidth]{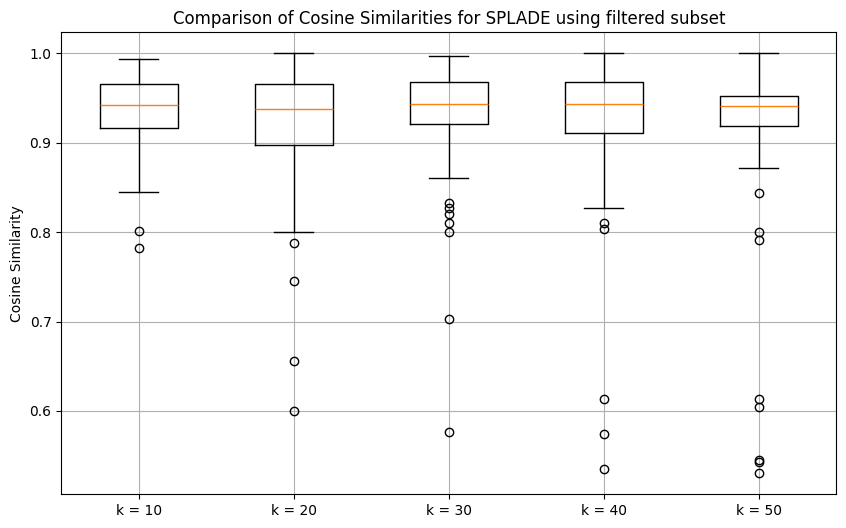}}
\caption{Boxplot of Cosine Similarities SPLADE}
\label{fig:SPLADE_COSSIM}
\end{figure}

\begin{figure}[htbp]
\centerline{\includegraphics[width=0.8\linewidth]{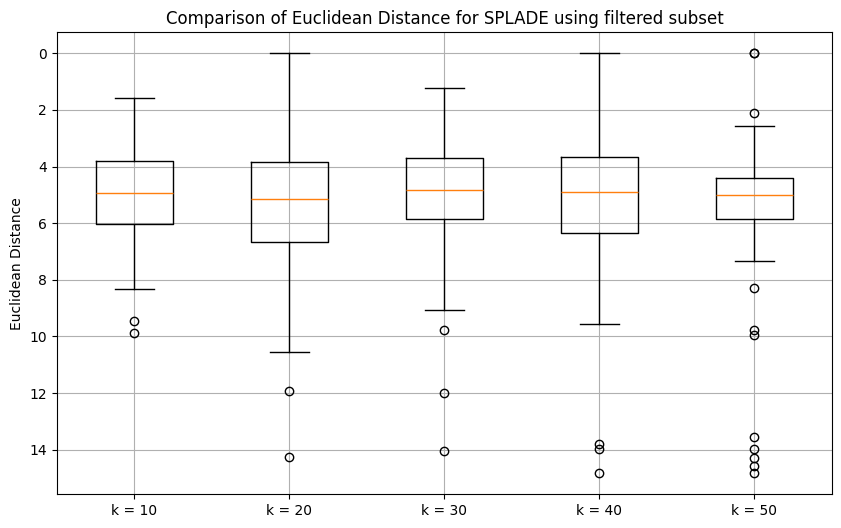}}
\caption{Boxplot of Euclidean Distances SPLADE}
\label{fig:SPLADE_EUCDIS}
\end{figure}

\end{document}